# Investigation of practical application for QAM hybrid receiver

Tian Chen,[1] Ke Li,[2] Yuan Zuo,[1] and Bing Zhu [1,*]

[1]Department of Electronic Engineering and Information Science, University of Science and Technology of China, Hefei, Anhui 230027, China
[2]Nanjing Research Institute of Electronic Technology, Nanjing, Jiangsu 230000, China
*Corresponding author: zbing@ustc.edu.cn

We present a quantum receiver for quadrature amplitude modulation (QAM) coherent states discrimination with homodyne-displacement hybrid structure. Our strategy is to carry out two successive measurements on parts of the quantum states. The homodyne result of the first measurement reveals partial information about the state and is forward to a displacement receiver, which finally identifies the input state by using feedback to adjust a reference field. Numerical simulation results show that for 16-QAM, the hybrid receiver could outperform the standard quantum limit (SQL) with a reduced number of codeword interval partitions and on-off detectors, which shows great potential toward implementing the practical application.

Coherent states are nonorthogonal to each other and then they cannot be discriminated without error [1]. However, coherent states have specially importance for communications since they are easy to prepare and manipulate and they are resilient to loss. This has led to significant effort to find and demonstrate measurement strategies for optimal discrimination of coherent states approaching the limits set by quantum mechanics [2-21]. The quantum mechanical bound of the minimum error rate is called Helstrom limit and is remarkably lower than the SQL which is attained by conventional receivers (direct detection, homodyne and heterodyne receiver) [1]. Quantum receivers to achieve the Helstrom limit were studied theoretically and then further investigated from a practical point of view for binary signals [2-6]. Experimentally demonstrations have been shown subsequently [7, 8]. Recently, Attention has been paid to multiple modulation signals [9-21]. Quantum receivers based on adaptive measurements [9-13] and classic-quantum hybrid structure [16] have also been proposed for M-ary phase shift keying (PSK) signals. A suboptimal receiver via the conditional pulse nulling (CPN) strategy is also demonstrated for M-ary pulse position modulation (PPM) signals [17, 18]. All of the above mentioned receivers contain two key parts, and they are displacement operation and photon counting, which can be implemented by beam splitters and single-photon detectors (on-off detectors or photon-number-resolving detectors: PNRD). The superiority of PNRDs to on-off detectors is verified resulting from the robustness against non-ideal devices [13-15].

Though people start to study the QAM signals discrimination [19, 20], the practical application of quantum receivers for QAM has not been reported which meets the need of both low cost and high signal repetition rate. Inspired by Müller [16] and Nair [21], we present a quantum receiver with hybrid structure consisting of a homodyne receiver and a subsequent displacement receiver, which reacts to the demand for practice.

The signals to be discriminated are M-ary QAM coherent states defined as,

$$|\varphi_{p,q}\rangle = |\alpha(p+jq)\rangle, \quad p,q \in \Omega \quad (1)$$

where $j=\sqrt{-1}$ and $\Omega = \{-(L-1)+2(i-1) | i=1,2,\cdots,L\}$. Without loss of generality, α can be taken as a real number. The number of signals M is represented by

$$M = L^2, \quad L=3,4,5,\cdots \quad (2)$$

Thus the average photon number $N_s$ is defined as,

$$N_s = \sum_{p,q} P_{p,q} |\alpha(p+jq)|^2 \quad (3)$$

where $P_{p,q}$ is a priori probability of QAM signals and through out this paper, we assume that $P_{p,q}=1/M$. The Helstrom limit for QAM signals discrimination is asymptotically given by the square root measurement (SRM) [22].

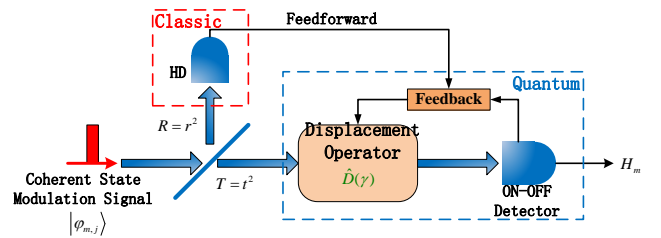

Fig. 1 Schematic of the quantum receiver with hybrid structrue.

The schematic diagram of the quantum receiver with hybrid structrue is shown in Fig. 1. The thick blue arrows indicate the optical signal and the thin black arrows indicate the electrical signal. The input signal is divided by a beam splitter (BS) with transmittance $T=t^2$ and reflectivity $R=r^2=1-T$. The transmitted and reflected parts are guided to a homodyne detector (HD) and a displacement-operator-based receiver, respectively. The local oscillating (LO) field is updated according to both the feedfoward and feedback results.

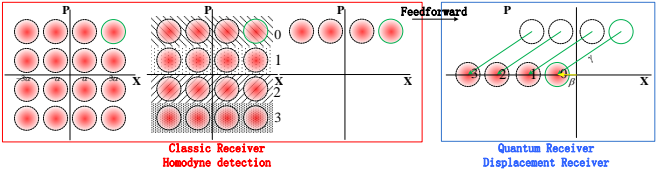

Fig. 2 Illustration of the measurements in phase space for 16-QAM signals.

As illustrated in Fig. 2, the HD detects along P quadrature in phase space and makes a decision on which row plane the signal is in. Then the result is forward to the displacement receiver, which is controlled by feedback strategy to find the column index of the remaining states and finally identifies the input state. So the homodyne-displacement (HD-D) hybrid receiver is much more efficient than other reported quantum receivers for QAM signals discrimination. It is also interested that such HD-D hybrid receiver has a natural advantage in QAM signals discrimination compared with PSK signals discrimination of equal M, especially when M is large.

The probability of success of individual classic and quantum receivers is independent, so the average error probability of hybrid receiver can be written as follows

$$P_E^{HD-D} = 1 - \sum_{p,q} P_{p,q} P_C^{HD_{p,q}} P_C^{D_{p,q}} \quad (4)$$

where $P_C^D$ and $P_C^{HD}$ represent the probability of correct detection of the displacement and the HD receiver, respectively.

We take 16-QAM signals discrimination for example.

The expression for correct detection corresponding to each row plane 0, 1, 2, 3 shown in Fig. 2 is

$$P_{00}^{HD}(\alpha) = erf\left(\sqrt{2}(p-3\alpha)\right)/2 \Big|_{p=2\alpha}^{p=+\infty}$$
$$P_{11}^{HD}(\alpha) = erf\left(\sqrt{2}(p-\alpha)\right)/2 \Big|_{p=0}^{p=2\alpha}$$
$$P_{22}^{HD}(\alpha) = erf\left(\sqrt{2}(p+\alpha)\right)/2 \Big|_{p=-2\alpha}^{p=0} \quad (5)$$
$$P_{33}^{HD}(\alpha) = erf\left(\sqrt{2}(p+3\alpha)\right)/2 \Big|_{p=-\infty}^{p=-2\alpha}$$

The displacement receiver sequentially nulls signals $0 \rightarrow 3 \rightarrow 1 \rightarrow 2$ (Type I) and uses an on-off detector to discriminate the four remaining hypotheses $H_{0,1,2,3}$. The probability of correct detection on each hypothesis is

$$P_{00}^D(\alpha) = p_0^N$$
$$P_{11}^D(\alpha) = p_1^{N-1}(1-p_1)/3 + \sum_{t=0}^{N-2} p_1^t(1-p_1) p_2^{N-2-t}(1-p_2)/2$$
$$+ \sum_{t=0}^{N-3} p_1^t(1-p_1) \sum_{s=0}^{N-3-t} p_2^s(1-p_2) p_0^{N-2-t-s}$$
$$P_{22}^D(\alpha) = p_2^{N-1}(1-p_2)/3 + \sum_{t=0}^{N-2} p_2^t(1-p_2) p_1^{N-2-t}(1-p_1)/2 \quad (6)$$
$$+ \sum_{t=0}^{N-3} p_2^t(1-p_2) \sum_{s=0}^{N-3-t} p_1^s(1-p_1) \sum_{u=0}^{N-3-t-s} p_1^u(1-p_1)$$
$$P_{33}^D(\alpha) = p_3^{N-1}(1-p_3)/3 + \sum_{t=0}^{N-2} p_3^t(1-p_3) p_0^{N-1-t}$$

where N is the number of partitions, i.e. the steps of feedback measurements, and

$$p_0 = \exp(-\nu), \quad p_1 = \exp(-\nu - 4\alpha^2\eta/N)$$
$$p_2 = \exp(-\nu - 16\alpha^2\eta/N), \quad p_3 = \exp(-\nu - 36\alpha^2\eta/N) \quad (7)$$

where $\nu$ and $\eta$ denote the dark count and quantum efficiency of detectors, respectively.

Consequently, the average error probability for the above HD-D hybrid receiver derived from equations (4), (5) and (6) is

$$P_E^{HD-D} = 1 - P_C^{HD-D}$$
$$= 1 - \left(\sum_{i=0}^{3} P_{ii}^{HD}(r\alpha)\right)\left(\sum_{j=0}^{3} P_{jj}^D(t\alpha)\right)/16 \quad (8)$$

Optimal parameters with N=10 for transmittance and displacement to minimize the average error rate are obtained in Fig. 3. The optimal transmittance T of BS for Type I receiver is shown in Fig. 3(a). It is clear that the optimal transmission parameter of the Type I receiver approaches to T=0.5. The error rate can also be reduced when the diaplacement ß is optimized. The optimal displacement parameters $|ß|^2$ with T=0.5 for the Type I and Type II (the displacement receiver sequentially nulls signals $0 \rightarrow 1 \rightarrow 2 \rightarrow 3$) receivers are shown in Fig. 3(b).

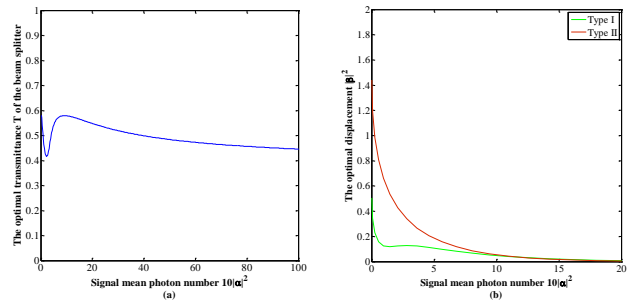

Fig. 3 Optimal parameters for (a) the transmittance T and (b) the displacement $|ß|^2$ with T=0.5 in the case of N=10.

Then we simulated error performances of different types of quantum receivers with ideal detectors for 16-QAM signals discrimination as shown in Fig. 4. The number of symbol partitions is set to be N=10 for comparision [20]. Meanwhile, the transmittance of BS is chosen to be T=0.5 where the near-optimal error probability is obtained, as shown in Fig. 3(a). Type III denotes the hybrid structure with adaptive measurements strategy. Type IV and Type V are adaptive measurements quantum receivers with on-off detectors and PNRDs, respectively. As indicated in the figure, Type IV receiver couldn't outperform the SQL. Besides, Type V receivers fail to beat the SQL at high photon number region resulting from the upward tendency of error rates as the signal mean photon number increases. However, all hybrid receivers can beat the SQL and the variation of the error rates with the signal mean photon number are almost monotonic. Error probabilities will be reduced when the optimal displacement (OD) or sequential probing order is adopted.

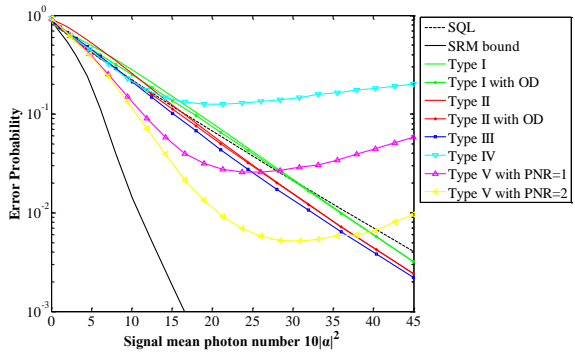

Fig. 4 Error performances of different types of quantum receivers for 16-QAM signals discrimination

Conclusively, a quantum receiver for QAM coherent states discrimination with HD-D hybrid structure is demonstrated theoretically. Numerical simulation results show that for 16-QAM, the hybrid receiver could outperform the SQL using a reduced number of symbol interval partitions compared with on-off-based adaptive measurements feedback quantum receiver. Its performance can be even better, especially at the weak-signal region when the diaplacement and transmittance parameters as well as the sequential probing order (feedback strategy) is optimized. In addition, the HD-D hybrid receiver is much more efficient to discriminate M-ary QAM signals, especially when M is large. It also needs to be pointed out that the HD-D hybrid receiver is potential with current technology for practical applications because few symbol partitions will allow us to detect the shorter pulsewidth or higher repetition rate signal and the usage of on-off detectors rather than PNRDs can significantly reduce the cost.

References:
[1] C. W. Helstrom, "Quantum detection and estimation theory," Academic Press, New York (1976).
[2] R. S. Kennedy, MIT RLE Quarterly Progress Report, Tech. Rep. 108, pp. 219-225 (1973).
[3] S. J. Dolinar, MIT RLE Quarterly Progress Report, Tech. Rep. 111, pp. 115-120 (1973).
[4] M. Takeoka, and M. Sasaki, Physical Review A **78**, 022320 (2008).
[5] V. A. Vilnrotter, NASA IPN Progress Report, vol. 42-189 (May 2012).
[6] D. Sych, and G. Leuchs, arXiv preprint arXiv: 1404.5033 (2014).
[7] K. Tsujino, D. Fukuda, G. Fujii, S. Inoue, M. Fujiwara, M. Takeoka, and M. Sasaki, Physical Review Letters **106**, 250503 (2011).
[8] R. L. Cook, P. J. Martin, and J. M. Geremia, Nature **446**, 774-777 (2007).
[9] R. S. Bondurant, Optics letters **18**, 1896-1898 (1993).
[10] F. Becerra, J. Fan, G. Baumgartner, S. Polyakov, J. Goldhar, J. Kosloski, and A. Migdall, Physical Review A **84**, 062324 (2011).
[11] S. Izumi, M. Takeoka, M. Fujiwara, N. Dalla Pozza, A. Assalini, K. Ema, and M. Sasaki, Physical Review A **86**, 042328 (2012).
[12] F. Becerra, J. Fan, G. Baumgartner, J. Goldhar, J. Kosloski, and A. Migdall, Nature Photonics **7**, 147-152 (2013).
[13] F. Becerra, J. Fan, and A. Migdall, Nature Photonics **9**, 48-53 (2015).
[14] S. Izumi, M. Takeoka, K. Ema, and M. Sasaki, Physical Review A **87**, 042328 (2013).
[15] K. Li, Y. Zuo, and B. Zhu, IEEE Photonics Technology Letters **25**, 2182-2184 (2013).
[16] C. R. Müller, M. Usuga, C. Wittmann, M. Takeoka, C. Marquardt, U. Andersen, and G. Leuchs, New Journal of Physics **14**, 083009 (2012).
[17] S. Dolinar Jr, The Telecommunications and Data Acquisition Progress Report, 42-72 (1983).
[18] J. Chen, J. L. Habif, Z. Dutton, R. Lazarus, and S. Guha, Nature Photonics **6**, 374-379 (2012).
[19] Y. Zuo, K. Li, and B. Zhu, arXiv preprint arXiv: 1412.4486 (2014).
[20] T.Chen, K. Li, Y. Zuo, and B. Zhu, arXiv preprint arXiv: 1504.02859 (2015).
[21] R. Nair, S. Guha, and S.-H. Tan, Physical Review A **89**, 032318 (2014).
[22] K. Kato, M. Osaki, M. Sasaki, and O. Hirota, Communications, IEEE Transactions on **47**, 248-254 (1999).